\documentclass[a4paper,11pt]{article}
\usepackage{pos}

\newcommand{\Pom}{\mathbb{P}}

\title{Exclusive $pp \to pp K^{*0} \bar{K}^{*0}$ reaction: $f_{2}(1950)$ resonance versus diffractive continuum}

\author{Piotr Lebiedowicz}

\affiliation{Institute of Nuclear Physics Polish Academy of Sciences\\
Radzikowskiego 152, PL-31342 Krak{\'o}w, Poland}

\emailAdd{Piotr.Lebiedowicz@ifj.edu.pl}

\abstract{
The exclusive reaction 
$pp \to pp (K^{*0} \bar{K}^{*0} \to K^{+}\pi^{-}K^{-}\pi^{+})$ 
for the LHC experiments is discussed. 
The amplitudes for the reaction are formulated 
within the tensor-pomeron approach.
We consider two diffractive mechanisms:
the $f_{2}(1950)$ $s$-channel exchange
mechanism and the $K^{*0}$-exchange mechanism.
First mechanism is a candidate 
for central diffractive production of tensor glueball 
and the second one is an irreducible continuum.
Comparison with data from WA102 experiment are made
and predictions for LHC experiments are given.
We find that including the continuum contribution alone 
one can describe the WA102 data reasonably well. 
A similar behaviour of the continuum and resonance contributions 
makes an identification of a broad tensor-glueball state 
in this reaction rather difficult.}

\FullConference{%
  *** The European Physical Society Conference on High Energy Physics (EPS-HEP2021), ***\\
  *** 26-30 July 2021 ***\\
  *** Online conference, jointly organized by Universität Hamburg and the research center DESY ***
}

\begin{document}
\maketitle

\vspace{-0.6cm}
\section{Formalism}
\begin{figure}
\begin{center}  
(a)\includegraphics[width=5.8cm]{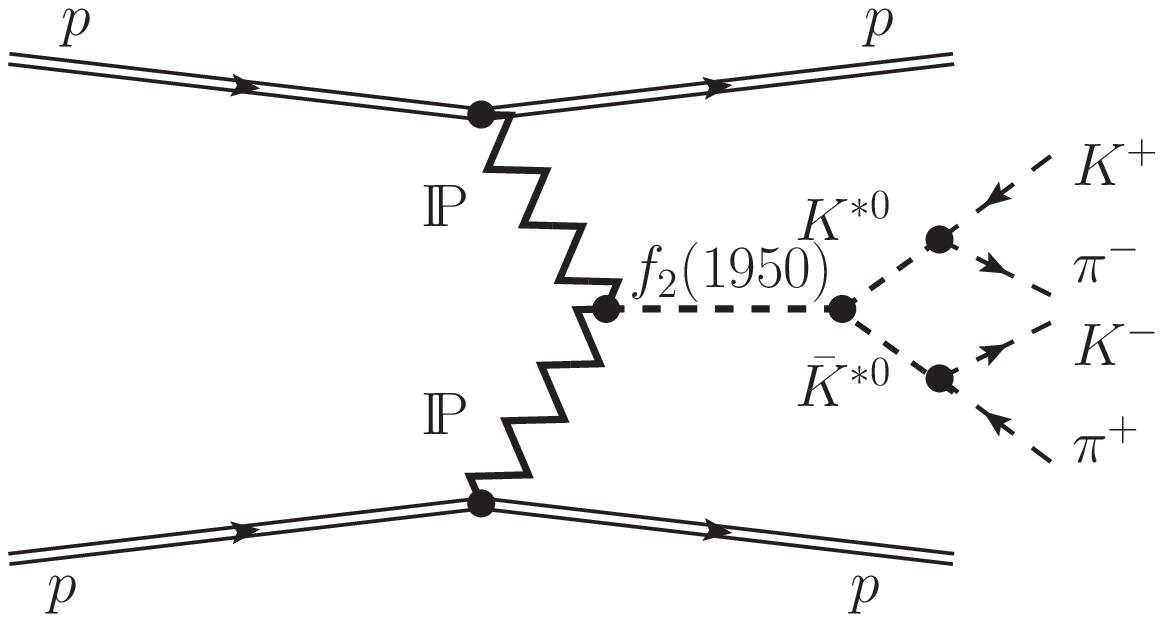} 
(b)\includegraphics[width=4.8cm]{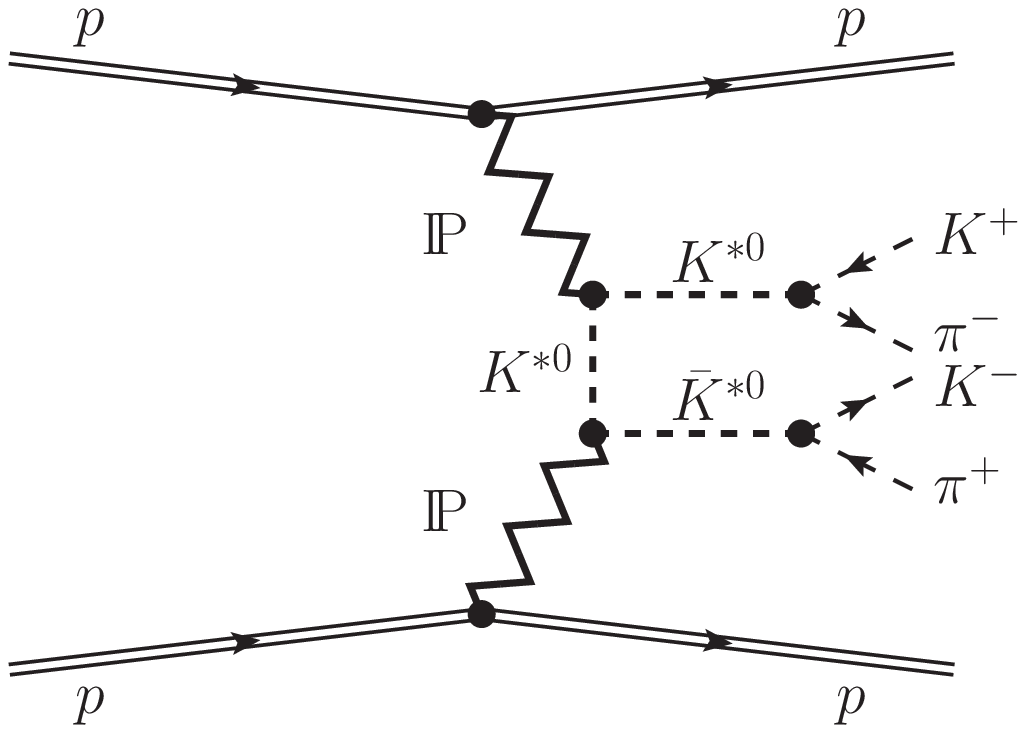}\\ 
\end{center}    
\caption{The ``Born-level'' diagrams for
CEP of $K^{*0} \bar{K}^{*0} (\to K^{+}\pi^{-} K^{-}\pi^{+})$ state in proton-proton collisions:
(a) the $\Pom \Pom \to f_{2}(1950)$ fusion mechanism,
(b) the continuum mechanism.}
\label{fig:diagrams}
\end{figure}

\vspace{-0.2cm}
In this contribution we discuss central exclusive production (CEP) of $K^{*0} \bar{K}^{*0} (\to K^{+}\pi^{-} K^{-}\pi^{+})$ state in proton-proton collisions. At high energies the pomeron-pomeron ($\Pom \Pom$) fusion processes (Figure~\ref{fig:diagrams}) are expected to be dominant.
We treat the $2 \to 6$ reaction effectively
as arising from the $2 \to 4$ reaction
with the spectral functions of $K^{*0}$ mesons.
We include absorptive corrections to the Born amplitudes 
in the one-channel eikonal approximation.
The presentation is based on \cite{1}
where all details and many more results can be found.
We treat the reaction in the tensor-pomeron approach \cite{2}.
There are many successful applications of 
the model 
to CEP reactions; see e.g. \cite{3,3b,3c,3d}.

The Born-level amplitude 
for the $2 \to 4$ reaction
$p(p_{a},\lambda_{a}) + p(p_{b},\lambda_{b}) \to
p(p_{1},\lambda_{1}) + p(p_{2},\lambda_{2}) + 
K^{*0}(p_{3},\lambda_{3}) + \bar{K}^{*0}(p_{4},\lambda_{4})$
via the $\Pom \Pom \to f_{2}(1950)$
fusion mechanism (Fig.~\ref{fig:diagrams}~(a))
can be written as
\begin{eqnarray}
&&{\cal M}^{(\Pom \Pom \to f_{2} \to K^{*}\bar{K}^{*})}_{\lambda_{a} \lambda_{b}
\to\lambda_{1}\lambda_{2} \lambda_{3}\lambda_{4}} 
= (-i)\,
\left(\epsilon^{(K^{*})}_{\kappa_{3}}(\lambda_{3})\right)^*
\left(\epsilon^{(\bar{K}^{*})}_{\kappa_{4}}(\lambda_{4})\right)^*
\bar{u}(p_{1}, \lambda_{1}) 
i\Gamma^{(\Pom pp)\,\mu_{1} \nu_{1}}(p_{1},p_{a})
u(p_{a}, \lambda_{a})\nonumber \\
&& \qquad \qquad \times 
i\Delta^{(\Pom)}_{\mu_{1} \nu_{1}, \alpha_{1} \beta_{1}}(s_{1},t_{1}) 
i\Gamma^{(\Pom \Pom f_{2})\,\alpha_{1} \beta_{1},\alpha_{2} \beta_{2}, \rho \sigma}(q_{1},q_{2}) \;
i\Delta^{(f_{2})}_{\rho \sigma, \alpha \beta}(p_{34})\;
i\Gamma^{(f_{2} K^{*}\bar{K}^{*})\,\alpha \beta \kappa_{3} \kappa_{4}}(p_{3},p_{4}) \nonumber \\
&& \qquad \qquad \times 
i\Delta^{(\Pom)}_{\alpha_{2} \beta_{2}, \mu_{2} \nu_{2}}(s_{2},t_{2}) \;
\bar{u}(p_{2}, \lambda_{2}) 
i\Gamma^{(\Pom pp)\,\mu_{2} \nu_{2}}(p_{2},p_{b})
u(p_{b}, \lambda_{b}) \,,
\label{amplitude_f2_pomTpomT}
\end{eqnarray}
where
$s_{1} = (p_{1} + p_{3} + p_{4})^{2}$,
$s_{2} = (p_{2} + p_{3} + p_{4})^{2}$, 
$q_{1} = p_{a} - p_{1}$, 
$q_{2} = p_{b} - p_{2}$, 
$t_{1} = q_{1}^{2}$, $t_{2} = q_{2}^{2}$,
and
$p_{34} = q_{1} + q_{2}$.
The effective pomeron-proton vertex and the tensor-pomeron propagator are [2],
\begin{eqnarray}
&&i\Gamma_{\mu \nu}^{(\Pom pp)}(p',p)
=-i 3 \beta_{\Pom NN} F_{1}(t)
\left\lbrace 
\frac{1}{2} 
\left[ \gamma_{\mu}(p'+p)_{\nu} 
     + \gamma_{\nu}(p'+p)_{\mu} \right]
- \frac{1}{4} g_{\mu \nu} ( p\!\!\!/' + p\!\!\!/ )
\right\rbrace \,,
\label{A4}\\
&&i \Delta^{(\Pom)}_{\mu \nu, \kappa \lambda}(s,t) =
\frac{1}{4s} \left( g_{\mu \kappa} g_{\nu \lambda} 
                  + g_{\mu \lambda} g_{\nu \kappa}
                  - \frac{1}{2} g_{\mu \nu} g_{\kappa \lambda} \right)
(-i s \alpha'_{\Pom})^{\alpha_{\Pom}(t)-1}\,,
\label{A1}
\end{eqnarray}
where $\beta_{\Pom NN} = 1.87$~GeV$^{-1}$,
$F_{1}(t)$ is the Dirac form factor of the proton, 
and $\alpha_{\Pom}(t)$ the pomeron trajectory:
$\alpha_{\Pom}(t) = \alpha_{\Pom}(0)+\alpha'_{\Pom}\,t$,
$\alpha_{\Pom}(0) = 1.0808$,
$\alpha'_{\Pom} = 0.25 \; {\rm GeV}^{-2}$ \cite{4a}.
A possible choice for the 
$i\Gamma_{\mu \nu,\kappa \lambda,\rho \sigma}^{(\Pom \Pom f_{2})(j)}$
coupling terms $j = 1, ..., 7$,
derived from a corresponding coupling Lagrangians,
is given in \cite{3,3b}. In this work we assume, that only the $j=1$ coupling,
corresponding to the lowest values of orbital angular momentum and spin of the two ``real pomerons'' $(l,S) = (0,2)$, is unequal to zero.
The $\Pom \Pom f_{2}$ vertex supplemented by form factors is
\begin{eqnarray}
i\Gamma_{\mu \nu,\kappa \lambda,\rho \sigma}^{(\Pom \Pom f_{2})} (q_{1},q_{2}) &=&
i\Gamma_{\mu \nu,\kappa \lambda,\rho \sigma}^{(\Pom \Pom f_{2})(1)}
\tilde{F}_{M}(q_{1}^{2}) \tilde{F}_{M}(q_{2}^{2}) F^{(\Pom \Pom f_{2})}(p_{34}^{2}) \nonumber \\
&=&
i\Gamma_{\mu \nu,\kappa \lambda,\rho \sigma}^{(\Pom \Pom f_{2})(1)}
\frac{1}{1-t_{1}/\tilde{\Lambda}_{0}^{2}}\;
\frac{1}{1-t_{2}/\tilde{\Lambda}_{0}^{2}}\;
\frac{\Lambda_{f_{2},P}^4}{\Lambda_{f_{2},P}^4 + (p_{34}^{2} - m_{f_{2}}^2)^{2}}
\,.
\label{vertex_pompomT}
\end{eqnarray}
The $\Pom \Pom f_{2}$ coupling constant ($g^{(1)}_{\Pom \Pom f_{2}}$) and form-factor cutoff parameters
($\tilde{\Lambda}_{0}$, $\Lambda_{f_{2},P}$) 
are treated as free parameters
which could be adjusted to fit the experimental data.
We take a simple Breit-Wigner form for the $f_{2}(1950)$-meson propagator.
The $f_{2} K^{*}\bar{K}^{*}$ vertex is as follows 
($M_{0} \equiv 1$~GeV):
\begin{eqnarray}
i\Gamma^{(f_{2} K^{*}\bar{K}^{*})}_{\mu \nu \kappa \lambda}(p_{3},p_{4}) =
i \left[\frac{2g'_{f_{2} K^{*}\bar{K}^{*}}}{M_{0}^{3}}  
\Gamma^{(0)}_{\mu \nu \kappa \lambda}(p_{3},p_{4})
F'(p_{34}^{2})- \frac{g''_{f_{2} K^{*}\bar{K}^{*}}}{M_{0}} \Gamma^{(2)}_{\mu \nu \kappa \lambda}(p_{3},p_{4})
F''(p_{34}^{2}) \right]\,,
\label{vertex_f2KKbar}
\end{eqnarray}  
with two rank-four tensor functions (see Eqs.~(3.18) and (3.19) of \cite{2}).
Here we assume
$F'^{(f_{2} K^{*}\bar{K}^{*})} = 
F''^{(f_{2} K^{*}\bar{K}^{*})} = 
F^{(\Pom \Pom f_{2})}$
and $\Lambda'_{f_{2}} = \Lambda''_{f_{2}} = \Lambda_{f_{2},P}$.
Thus, the result depends on $\Lambda_{f_{2},P}$ and 
the product of the couplings
$g^{(1)}_{\Pom \Pom f_{2}} g'_{f_{2} K^{*}\bar{K}^{*}}$ or
$g^{(1)}_{\Pom \Pom f_{2}} g''_{f_{2} K^{*}\bar{K}^{*}}$.
In the following we assume that only either
the first or the second of the product of couplings 
is nonzero.

In the continuum mechanism, we take into account 
the reggeization of intermediate $K^{*}$ meson; see~\cite{1}.
We used two parametrisations of the $K^{*}$ trajectory:
linear and nonlinear (square-root form)~\cite{5}.
The $\Pom K^{*}K^{*}$ vertex,
with $k', \mu$ and $k,\nu$ the momentum and vector index
of the outgoing and incoming~$K^{*}$, respectively,
and $\kappa \lambda$ the tensor $\Pom$ indices, reads
\begin{eqnarray}
i\Gamma^{(\Pom K^{*}K^{*})}_{\mu \nu \kappa \lambda}(k',k) =
i \left[
2a_{\Pom K^{*}K^{*}} 
\Gamma^{(0)}_{\mu \nu \kappa \lambda}(k',-k)
- b_{\Pom K^{*}K^{*}}\,\Gamma^{(2)}_{\mu \nu \kappa \lambda}(k',-k) \right] \frac{1}{1-(k'-k)^{2}/\Lambda_{0}^{2}}
\hat{F}_{K^{*}}(\hat{p}^{2}).
\label{vertex_pomKK}
\end{eqnarray}  
Here, the form factors $\hat{F}_{K^{*}}(\hat{p}_{t}^{2})$
and $\hat{F}_{K^{*}}(\hat{p}_{u}^{2})$
are parametrised in the exponential form.
We assume that $a \neq 0$, $b = 0$ 
or $a = 0$, $b \neq 0$.
The coupling and cutoff parameters ($a_{\Pom K^{*}K^{*}}$, $b_{\Pom K^{*}K^{*}}$, $\Lambda_{0}$, $\Lambda_{\rm off,E}$) 
could be adjusted to experimental data; see Ref.~\cite{1} for their numerical values and for more details.

\vspace{-0.2cm}
\section{Comparison with the WA102 data and predictions for the LHC experiments}
\begin{figure}
\begin{center}
\includegraphics[width=0.32\textwidth]{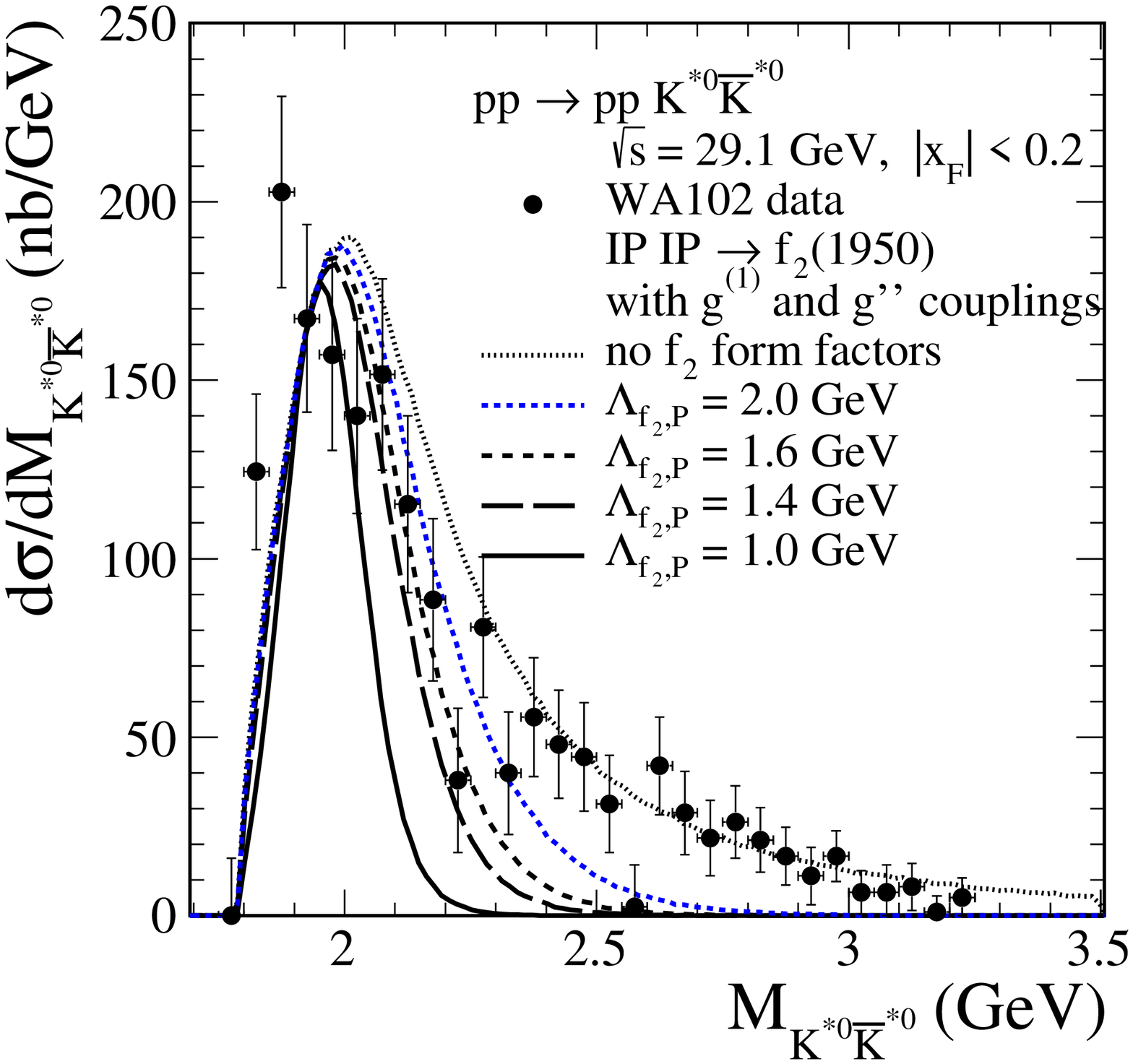} 
\includegraphics[width=0.32\textwidth]{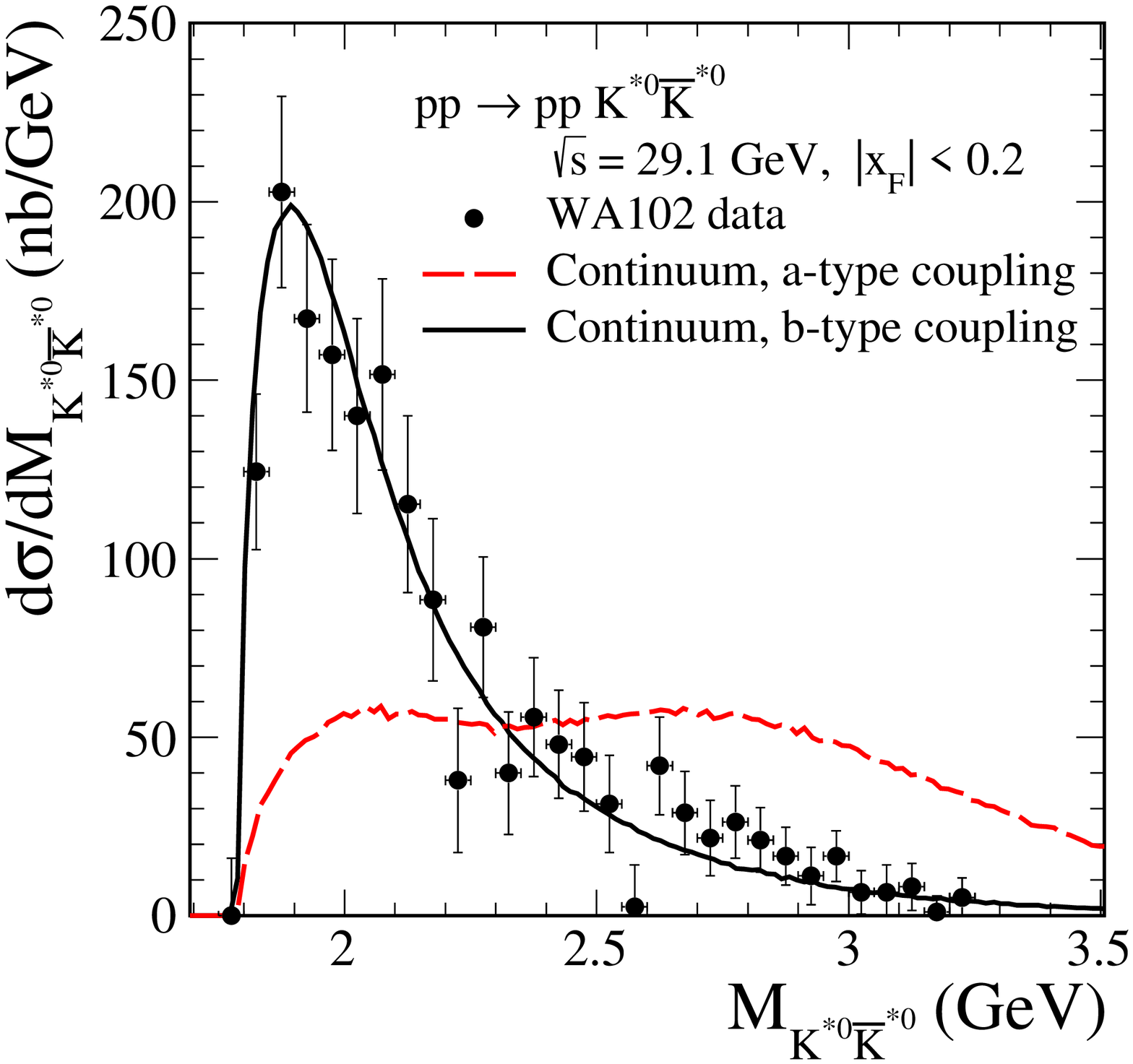}
\includegraphics[width=0.32\textwidth]{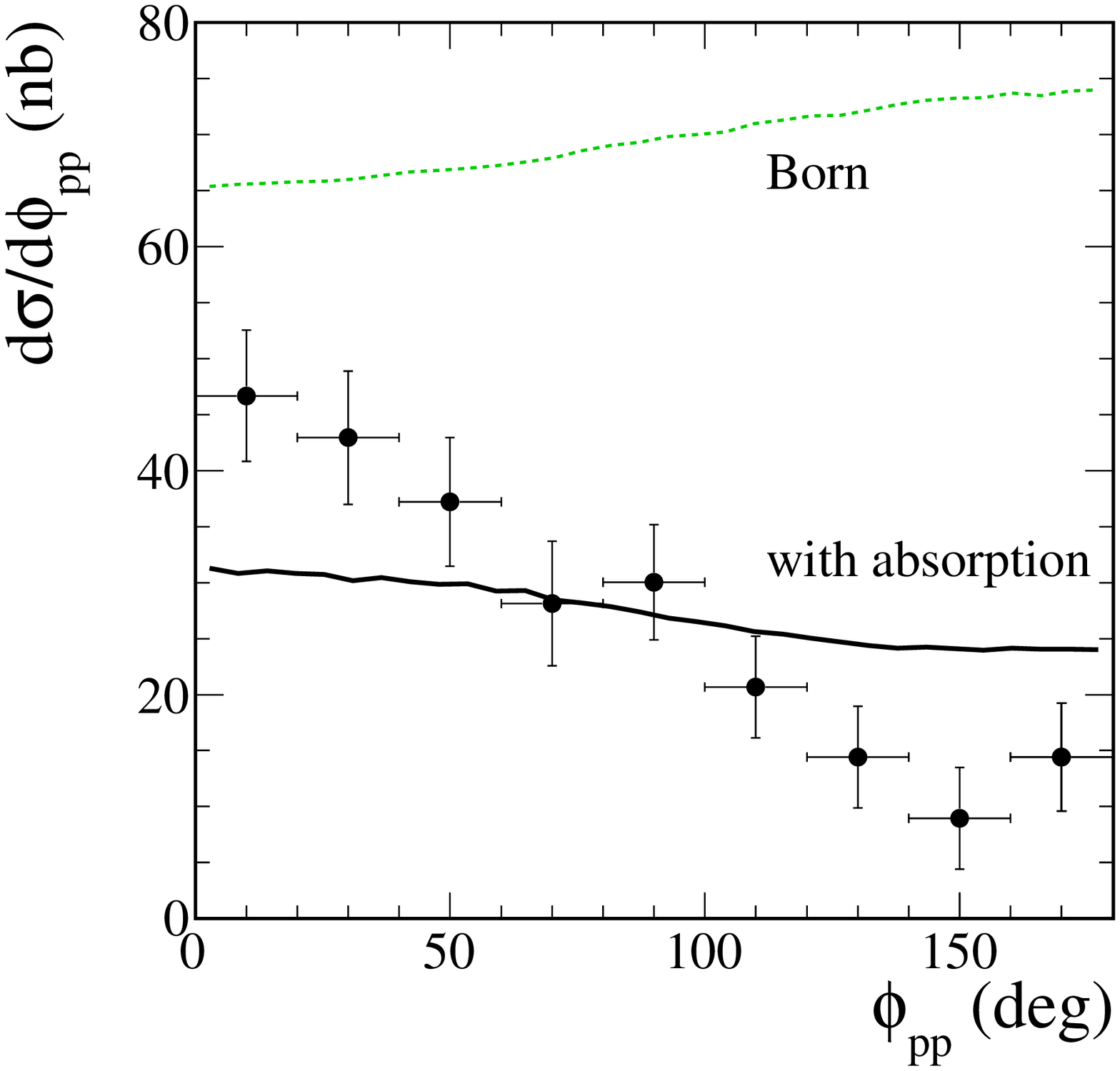}
\end{center}  
\caption{The distributions in $K^{*0} \bar{K}^{*0}$ invariant mass
for the $f_{2}(1950)$ mechanism (left),
for the continuum mechanism (center),
and the $\phi_{pp}$ distribution (right)  
together with the WA102 data ($\sigma_{\rm exp} = 85 \pm 10~{\rm nb}$ \cite{4}).}
\label{fig:WA102}
\end{figure}

\begin{figure}
\begin{center}
\includegraphics[width=0.32\textwidth]{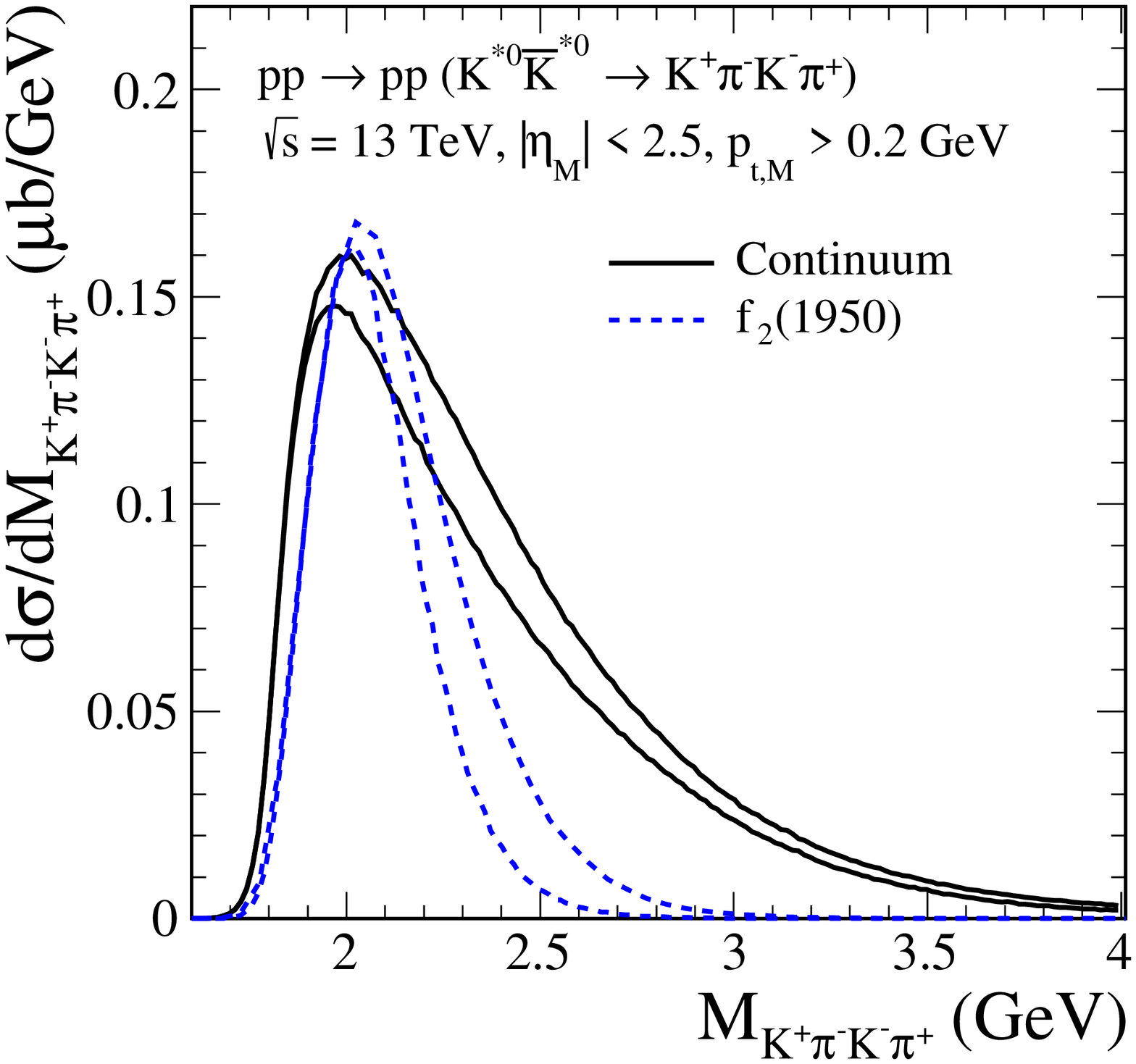}
\includegraphics[width=0.32\textwidth]{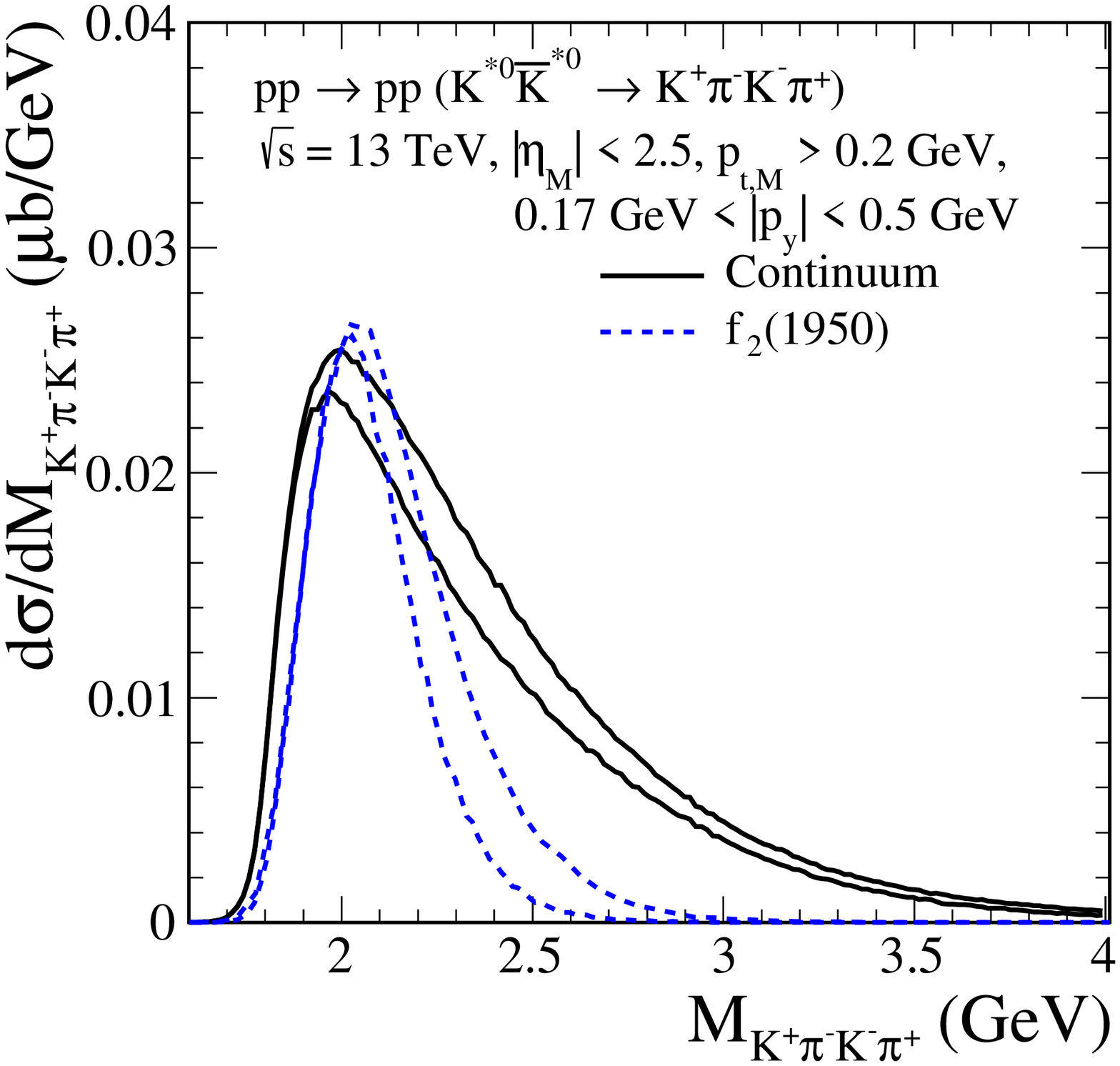}
\includegraphics[width=0.32\textwidth]{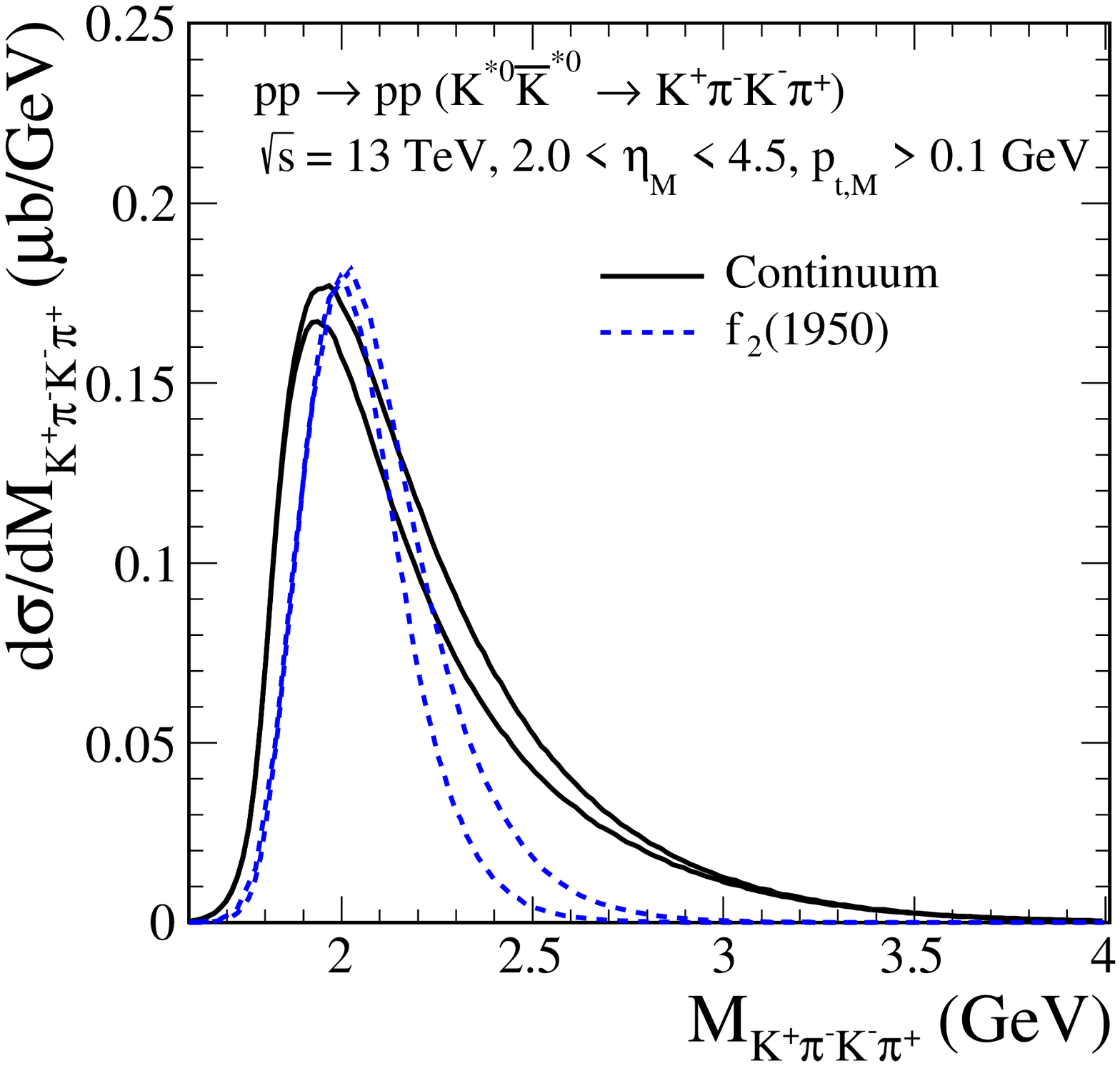}\\
\end{center}  
\caption{Invariant mass distributions for the $K^{+}\pi^{-}K^{-}\pi^{+}$ system
calculated for $\sqrt{s} = 13~{\rm TeV}$.
We show the results with cuts on 
pseudorapidities and transverse momenta of produced pions and kaons, 
and with an extra cuts
on momenta of outgoing protons that will be
measured in the ATLAS+ALFA experiment (in the middle panel),
and the results for larger $\eta_{M}$
and without a measurement of protons
relevant for the LHCb experiment (right panel).
For the continuum term we show the results 
for two parametrisations of the $K^{*}$ trajectory:
the linear form (lower solid line) and
the nonlinear form (upper solid line).
For the $f_{2}$ contribution the results for 
$\Lambda_{f_{2},P}$ = 1.6~GeV (lower dotted line), 
2~GeV (upper dotted line), 
and with $|g^{(1)}_{\Pom \Pom f_{2}} g''_{f_{2} K^{*}\bar{K}^{*}}|=11$
are presented.}
\label{fig:LHC}
\end{figure}

\vspace{-0.2cm}
In our exploratory study we consider separately 
the two mechanisms shown by the diagrams in Fig.~\ref{fig:diagrams}.
We obtain a good description of the WA102 data \cite{4}
for the reaction $pp \to pp K^{*0} \bar{K}^{*0}$
assuming the dominance of pomeron-pomeron fusion
already at $\sqrt{s} = 29.1~{\rm GeV}$ (Fig.~\ref{fig:WA102}).
The model results, in both cases,
are in better agreement with the WA102 data 
for the tensor-vector-vector coupling vertices
$\propto \Gamma^{(2)}$.
The absorption effects were included.
Absorption effects lead to a reduction of the cross section
and change the shape of the $\phi_{pp}$ distribution, the azimuthal angle between 
the transverse momentum vectors of the outgoing protons.
Both considered mechanisms 
have a maximum around $M_{2K2\pi} \simeq 2$~GeV (see Fig.~\ref{fig:LHC}),
thus a broad enhancement
in this mass region 
can be misidentified as the $f_{2}(1950)$ resonance
(one of the tensor glueball candidates).
The predictions for the LHC experiments 
should be regarded rather as an upper limit.

\vspace{-0.2cm}
\section{Conclusions}
\vspace{-0.2cm}
\begin{itemize}
\item 
The calculation for the $pp \to pp K^{*0} \bar{K}^{*0}$ reaction have been performed in the tensor-pomeron approach \cite{2}. We have discussed
CEP of the $f_{2}(1950)$ resonance
and the continuum with the intermediate $K^{*}$-reggeized exchange.
We obtain a good description of the WA102 data
with the continuum contribution alone,
assuming that the reaction is dominated by pomeron-pomeron fusion.
\vspace{-0.2cm}
\item
Predictions for the reaction $pp \to pp K^{+} K^{-} \pi^{+} \pi^{-}$ for the LHC experiments at
$\sqrt{s} = 13~{\rm TeV}$ were given.
We obtain $\sigma \simeq 17-250~{\rm nb}$
depending on the assumed cuts.
Absorption effects were included.
Similar behaviour of considered mechanisms 
(Fig.~\ref{fig:diagrams})
makes an identification of a broad tensor-glueball state in this reaction rather difficult.
\end{itemize}

\end{document}